\documentstyle[12pt,aasms4]{article}
\def\day{{}$^{\rm d}$\llap{.}}
\accepted{ July 30, 2001}

\begin{document}

\title{VARIABLE STARS IN GALACTIC GLOBULAR CLUSTERS}

\author{Christine M. Clement, Adam Muzzin, Quentin Dufton, Thivya Ponnampalam, 
John Wang,
Jay Burford, Alan Richardson and Tara Rosebery}

\affil{Department of Astronomy and Astrophysics, University of Toronto \\
Toronto, ON, M5S 3H8, CANADA\\
electronic mail: cclement@astro.utoronto.ca,muzzin@ungrad.astro.utoronto.ca,
dufton@physics.utoronto.ca,thivya.ponnampalam@utoronto.ca,
johnny.wang@utoronto.ca,jburford@home.com,al.richardson@utoronto.ca}

\author{Jason Rowe}

\affil{Department of Physics and Astronomy, UBC \\
Vancouver, BC, V6T 1Z4, CANADA\\
electronic mail: rowe@astro.ubc.ca}

\and

\author{Helen Sawyer Hogg
\altaffilmark{1}}
\altaffiltext{1}{obiit}

\begin{abstract}
Based on a search of the literature up to May 2001,
the number of known variable stars in Galactic globular clusters is 
approximately 3000.
Of these, more than 2200 have known periods and 
the majority (approximately 1800) are of the RR Lyrae type. 
In addition to the RR Lyrae population, there are approximately 100
eclipsing binaries, 120 SX Phe
variables, 60 Cepheids (including population II Cepheids, anomalous Cepheids 
and RV Tauri) and 120 SR/red variables.
The mean period of the fundamental mode RR Lyrae variables is 0\day 585,
for the overtone variables it is 0\day  342 (0\day 349 for the 
first-overtone pulsators and 0\day 296 for the second-overtone pulsators)
and approximately 30\% are overtone pulsators.
These numbers indicate that about 65\%
of RR Lyrae variables in Galactic
globular clusters belong to Oosterhoff type I systems. The mean period of
the RR Lyrae variables in the Oosterhoff type I clusters seems to be
correlated with metal abundance in the sense that the periods are longer
in the more metal poor clusters. Such a correlation does not exist for the
Oosterhoff type II clusters.
Most of the Cepheids are in clusters with blue horizontal branches.
\end{abstract} 


%
%
\section{INTRODUCTION}

Variable stars are useful
standard candles for determining the distances to the Galactic globular
clusters (GGCs) and it is important to know these
distances for an understanding of
the age, structure and formation of the Galaxy. 
As a result, there are
many papers on the subject of variable stars in globular clusters.
More than sixty years ago, Sawyer (1939) published a catalog of variable
stars in GGCs
to enable investigators interested in the topic to
get a clear picture of exactly what work had been done in the area.
The catalog listed 1116 variables in 60 clusters. 
Second and third editions of the catalog, with 1421 and 2119 entries
respectively, were subsequently produced (Sawyer 1955, Sawyer Hogg 1973).
Sawyer Hogg
intended to publish a fourth edition, and in preparation for this,
she recorded material from the relevant papers published
between 1973 and 1988 on reference cards. 
After her death in 1993, the cards were placed with the 
University of Toronto Archives and Records Management Services (UTARMS) where
they now reside.
In 1997, an electronic version of
the 1973 catalog was produced and the material from the 1973--1988
reference cards was included (Clement 1997).

Since 1988, there have been many 
papers on the subject of
globular clusters and their variables.  Furthermore, during
this period CCD detectors have been widely used. 
Consequently, many new variables have been discovered, particularly in
crowded central regions of clusters.
Thus it is an appropriate time to update the catalog and
prepare a summary of the material.
To search the literature for papers published after 1988, we
consulted volumes 49-58 and 61-68 of the Astronomy and Astrophysics Abstracts
for the years 1989-1993 and 1994-1997 respectively, and for papers 
published after 1997, we consulted the 
NASA Astrophysics Data System. In this paper,
we summarize the numbers and types of variable stars in
GGCs. In addition, for the  
RR Lyrae variables and Cepheids, we illustrate how the periods 
relate to the clusters' metal
abundance and horizontal branch morphology.  The complete updated
catalog can be obtained at the following website: 

http://www.astro.utoronto.ca/people.html

%
%

\section{SUMMARY OF THE DATA}

In Table 1, we list the 
Galactic globular
clusters which are known to contain variables. For each cluster,
we list the horizontal branch
ratio (HBR) and the metal abundance, [Fe/H]. The HB ratio, $(B-R)/(B+V+R)$, 
is a parameter devised
by Lee (1990) to describe horizontal branch morphology. Both of
these quantities
were taken from the 1999 update to Harris's (1996) catalog of globular cluster
parameters ($\rm{http://www.physun.physics.mcmaster.ca/Globular.html}$).
The total number of variables in each cluster
and the number for which periods have
been determined is listed in column (3) and the main types of variables are
indicated in columns (4) to (9). These totals do not include any stars that
are suspected or confirmed to be field stars. 
`SX' (column 4) refers to the SX Phe type variables, stars in the region just
above the ZAMS, that pulsate with periods less than 0\day 2.
`RR' (column 5) refers to RR Lyrae variables.  Although many 
variables have been 
tentatively classified as RR Lyrae, we include here only the ones for 
which periods have been determined.
`Cep' and `RV Tau' (column 6)
refer to Cepheids (anomalous or population II) and RV Tauri variables.
In general, we have classified 
stars with periods less than 1 day as RR Lyrae variables and stars with
periods greater than 1 day as Cepheids. 
However, the actual threshold
between these two groups of variables may be P=0\day 75 or 0\day 80
(Wallerstein \& Cox 1984, hereafter WC84; Gautschy \& Saio 1996, hereafter 
GS96).
The `SR/Red' variables (column 7) are pulsating variables
that have periods greater than 35 days and/or stars
that have been classified as semi-regular
or irregular. 
Column (8) lists the number of eclipsing binaries and in 
column (9), we cite the papers in which new information has been presented
concerning the elements of the variables  since Sawyer Hogg's (1973) 
catalog. For most of the clusters, the totals of columns 4 to 8 are equal to
or greater than the number of variables for which periods have been
determined because many irregular variables do not have published periods.
However, in some cases (e.~g. M15), there are `unclassified' variables with
published periods.

There are 147 clusters in the Harris catalog, but only 102 are known to
have variable stars. This does not mean that there are none
in the other 45 because not all clusters have been 
searched. Also,
with the use of CCD detectors  and new search techniques, for example,
the image subtraction method introduced by  Alard \& Lupton (1998),
many new variables will be discovered,
even in the clusters that have already been studied.

In Figure 1, we plot the period-frequency
distributions for pulsating variables in globular clusters.
In the top panel, stars with periods less than 1 day are plotted.
The stars with P$<$0\day 2 are considered to be SX Phe variables. Most of
the others are RR Lyrae variables, but a few of the longer period ones
may be anomalous Cepheids or population II Cepheids. The middle panel
includes the Cepheid and RV Tauri variables with periods in the range 1 to 
50 days. The RV Tauri stars are plotted with their pulsation
periods. The plot also includes two 
stars (one in $\omega$ Centauri and one in NGC 5466) with periods 
less than 1 day and classified as anomalous Cepheids. 
In the lower panel, variables
with pulsation periods between 35 and 440 days are shown.

Figure 2 shows the period-frequency
distribution for the eclipsing binaries.
The majority of these stars have periods less than 1 day and about
half are of the W Ursa Majoris type. 
Rucinski (2000) recently reviewed the
literature on W UMa-type binaries and found that at least one-third of
his sample were
foreground stars. We have not included any of these foreground stars in
Figure 2 or in Table 1.

%
%
\section{THE SX PHE VARIABLES}

Rodr\'iguez \& L\'opez-Gonz\'alez (2000) published
a catalog of SX Phe stars in GGCs and in nearby 
galaxies. They showed
that there seems to be a correlation between mean periods 
and metallicity in the sense that the 
periods of the SX Phe variables are longer in stellar systems with
higher metallicity. They also pointed out that all of the SX Phe
variables  have been discovered in the last 20 years and that there has
been a great increase in their numbers
during the last few years. As more clusters
are studied, the number of SX Phe stars
will undoubtedly increase further and we will be able to gain a better
understanding of  how their properties relate to the properties of the
clusters to which they belong. An interesting point to note from the
data of Table 1 is that, in
four of the 21 clusters known to have SX Phe stars, no RR Lyrae stars
have been detected. Three of these, NGC 4372, 6397 and 6752, have
extremely blue horizontal branches and the other one, M71, has an
extremely red HB. Since SX Phe stars are near the main sequence, it seems 
that  they are not affected by horizontal branch morphology.

%
%

\section{THE RR LYRAE VARIABLES}

For the RR Lyrae variables, we have adopted a system of notation
recently introduced by the MACHO consortium (Alcock et al. 2000, hereafter
A00). `RR0' instead of `RRab' is
used to designate fundamental mode pulsation, `RR1' instead of `RRc'
for first-overtone,
`RR01' instead of `RRd' for double-mode (fundamental and first-overtone), 
`RR2' instead of `RRe' for second-overtone pulsation and `RR12' for
double-mode (first and second-overtone). No RR12 variables
have yet been identified
in globular clusters, but A00 found three in the LMC.
In Table 2, we list information concerning the periods of
RR Lyrae variables in the GGCs:
in columns (2) and (3), the number of fundamental
mode pulsators and their mean periods, in columns (4) and (5), the number
of first overtone pulsators and their mean periods, in column (6) and (7),
the number of second overtone pulsators and their mean periods, in columns
(8) and (9), the number of double-mode pulsators and their mean 
periods and in column (10), the mean fundamentalized period.
Since  the dominant mode of pulsation for most of the double-mode
variables is the first-overtone, we list the mean overtone 
periods in column (9). The fundamentalized period was computed
assuming period ratios: $P_1/P_0=0.745$ and $P_2/P_1=0.804$, typical
period ratios derived by A00 for the RR01 and RR12
variables in the LMC.
The possibilty that RR2 variables really exist
has been the subject of some
discussion in the literature. In recent studies of $\omega$
Cen and NGC 5897, Clement \& Rowe (2000, 2001) showed that the
RR1 and RR2 variables seem to separate into two sequences in the
period-luminosity and period-amplitude plots. 
Therefore, to decide whether an overtone pulsator should be classified as
RR1 or RR2, we consulted the period-amplitude relation of
Clement \& Rowe (2000). 
However, it is not always easy to discriminate between RR1
and RR2 variables in the period-amplitude plot because the models of 
Bono et al.
(1997) predict that the period-amplitude relation for RR1 stars shows
a characteristic `bell' shape with amplitudes decreasing at shorter
periods. 
Thus in some cases, our classifications are uncertain.

According to Table 2, the mean period for all of the RR0 variables in GGCs
is 0\day 585 and for the overtone variables, it is 0\day 342 (0\day 349
for the first-overtone and 0\day 296 for the second-overtone). 
Approximately 30\% of the RR Lyrae variables are overtone pulsators.  
In calculating the mean overtone periods, we have included the RR01
variables as first-overtone pulsators because the first-overtone is
generally their dominant mode of pulsation. 
These figures
indicate that most of the GGC RR Lyrae variables are in Oosterhoff
type I systems.\footnote{ Oosterhoff (1939, 1944) recognized that globular
clusters could be classified into two groups
according to the period-frequency distribution of their RR Lyrae
variables. He showed that the mean periods for the RR0 variables 
were $\sim$0\day 55 in the type I (OoI) clusters and $\sim$0\day 65 in the
OoII clusters. He also found that, in the OoII clusters, the percentage
of overtone pulsators was higher than in the OoI clusters.} 
In their investigation of the MACHO data for LMC RR Lyrae variables,
Alcock et al. (1996, hereafter A96) reached a similar conclusion.

In Figure 3, we show the period-frequency distributions for the RR Lyrae
variables separated according to Oosterhoff type. In assigning the Oosterhoff
type, we assume that
all clusters for which the mean period of the RR0 stars is $<$0\day 60 
belong to group I and that clusters for which the mean RR0 period is
$\ge$0\day 60, with the exception of  Ruprecht 106, belong to group
II. The study of the RR Lyrae
variables in Rup 106 by Kaluzny et al. (1995) shows that its period-amplitude
relation is similar to that of M3 and so we have classified it as OoI. 
Pal 5, which has no RR0 stars, is considered to be OoI on the basis of 
the mean period of its RR1 variables. When the period-frequency data
are separated by Oosterhoff type, it is possible to see that there are two
peaks in the distribution
for the overtone pulsators, particularly in the OoII clusters. This is an 
effect that A96
noted in the MACHO LMC data for RR Lyrae variables and they attributed it to
pulsation in the second-overtone mode. The mean period for the
936 RR0 stars in the OoI clusters is 0\day 559, for the 232 RR1 and RR01
stars, it is 0\day 326, for the 29 RR2 stars, it is 0\day 281, the mean
fundamentalized period is 0\day 533,
and 22\% of the variables are overtone pulsators. The equivalent
figures for the OoII clusters are 0\day 659 for 333 RR0,
0\day 368 for 268 RR1 and RR01, 0\day 306 for 44 RR2,
0\day 580 for $<P_f>$ and 48\% are overtone 
pulsators. These numbers are comparable to what Oosterhoff (1939) found
for the two groups, but the percentage of the (lower amplitude)
overtone pulsators\footnote{When Oosterhoff discovered the two cluster
groups in 1939, he used the terms `c-type' and `a- and b-type' for 
first-overtone and fundamental-mode pulsators. This is because it was not 
recognized until the following year that the c-type variables
were pulsating in the first-overtone mode (Schwarzschild 1940).}
has increased.  

In Figures 4 to 7, we plot [Fe/H] and HBR versus the mean
periods of the RR0 and RR1\footnote{No plots were made for the
RR2 stars because there are not enough of them. $\omega$ Cen is the only
cluster in which we have classified more than 5 stars as RR2. We considered 
the possibility 
that our assumed mean periods for the RR1 stars would be in error
if the stars we considered to be RR2 variables were in fact RR1
variables. However, it turned out that such errors would not be large
enough to affect the appearance of Figures 6 and 7.}
variables.
Since any correlations are more significant
if the sample of RR Lyrae variables is greater, we plot in the
upper panels only the systems with at least 15 RR0 stars
(Figures 4 and 5) or at least 15
RR1 stars (Figures 6 and 7) and then in the lower panels we include systems
with fewer variables. Similar plots for the  mean `fundamentalized' periods
are shown in Figures 8 and 9.
In Figure 4, there seems to be
a correlation between [Fe/H] and mean period for the OoI clusters (i.e. the
ones with $<P0>$ less than 0\day 60) in
the sense that the periods are longer in the 
more metal poor clusters. 
However, the correlation breaks down when the mean period is
greater than 0\day 60, i.e. in the OoII clusters.
A similar trend can be seen for the mean fundamentalized periods in Figure 8. 
The mean period is correlated with [Fe/H] 
for clusters with $<P_f>$ less than about 0\day 56 (OoI clusters),
but not for clusters with longer periods.
The period-[Fe/H] correlation can also be seen among the RR1 mean
periods less than 0\day 34 in Figure 6, but it
is not as marked as in Figures 4 and 8 because
OoI clusters do not have so many RR1 variables. 

Until recently, it
has been assumed that metal rich clusters have RR Lyrae variables with
short periods and therefore belong to the OoI class,
but investigations of 
NGC 6388 and NGC 6441 by Pritzl et al. (2000) and Layden et al. (1999)
have shown that these two clusters are exceptions to
the rule. Both have  [Fe/H]$\sim -0.6$ which makes them the
most metal rich clusters with RR Lyrae variables, but their
$<P0>$ values are greater than 0\day 70. Because of these long periods,
we classify these clusters as OoII.
The discrepant `crosses' in the middle
and lower panels of Figure 4, the lower panel of Figure 6 and the middle
and lower panels of Figure 8 are the points for these clusters. In fact, 
NGC 6441 does not even appear in Figure 8.
Since it has no known RR1 stars, its mean fundamentalized period is
so long (0\day 768) that it is off the scale of the diagram.
The metal rich cluster 47 Tuc (NGC 104) also has an RR0 variable with a
period longer than 0\day 70 (Carney et al. 1993), but it does not appear
in Figures 4 or 8 because it is the only RR Lyrae variable in the
cluster.

The importance of [Fe/H] for determining properties of
globular cluster RR Lyrae variables has already been questioned
by Clement \& Shelton (1999, hereafter CS99) and by Lee \& Carney (1999b).
Lee \& Carney compared M2 and M3, two clusters with similar metal
abundance and found substantial differences in their RR Lyrae populations.
CS99 studied the period $V$-amplitude relation for RR0 stars
with `normal' light curves\footnote{The compatibility test of
Jurcsik \& Kovacs (1996) was used to assess whether or not
the light curve of a star was normal.}
in several globular clusters 
and found that the $V$ amplitude for a given period was a function of
Oosterhoff type, but not a function of [Fe/H]. 
The OoII clusters in their sample were M9 and M68, but they also included
two M92 RR0 stars for which published CCD photometry was available.
Subsequently, Clement (2000a) showed that the RR0 variables in
in the OoII cluster M55 and in
the metal rich clusters 47 Tuc and NGC 6441, as well as
the RR0 stars brighter than $V=14.65$ in 
$\omega$ Cen  (the OoII variables) 
all fit the OoII relation of CS99.  In the meantime,
Papadakis et al. (2000) confirmed  that the RR0 variables in the OoII 
cluster NGC 6426 fit the OoII period-amplitude relation
derived by CS99.  More recently, Kopacki (2001) compared the 
period-amplitude relations for the RR0 variables in three OoII clusters
with different [Fe/H],
M2, M53 and M92, and verified that there was no significant period-shift.
Thus period-shift for the OoII clusters is not a function of
metal abundance.  The P-A relation plotted by Alves et al. (2001)
for NGC 5986 also seems to coincide with the CS99 OoII relation.
An interesting feature of their diagram is that the P-A relation for
NGC 5986 is shifted to longer periods than the fiducual ridge line
they plot for M15. 
Since M15 is more metal poor than NGC 5986, they were expecting the M15 RR0
stars to have longer periods.
M15 which is often considered the prototype for
OoII clusters seems to have a P-A relation that is different from
the others! As noted above,
Figures 4, 6 and 8 also illustrate that the periods for OoII clusters do not
depend on [Fe/H].
It therefore seems well established that, for the OoII clusters, mean 
periods and period shifts are not correlated
with [Fe/H]. 

However, the situation is different in the OoI clusters.
We have seen from Figures 4 and 8 that there is a correlation
between mean period and [Fe/H] for the OoI clusters.
The OoI clusters that CS99 used for their study were M3 amd M107. 
They acknowledged that the transition 
between fundamental and first-overtone mode pulsation
occurs at a shorter period in M107 compared to 
M3. However, the short period fundamental mode pulsators in M107 did not
have `normal' light curves. As a result, there was no
period shift between the period-amplitude relations for `normal'
RR0 stars in the two clusters.
Later, Kaluzny et al. (2000) performed a similar analysis 
for the RR0 variables in M5 and found that its P-A relation 
was shifted to shorter periods compared with M3. Borissova et al. (2001)
reached the same conclusion in a recent study of NGC 6229. Thus
in the OoI clusters, mean periods and period shifts seem to be correlated
with [Fe/H], but in the OoII clusters they are not.
This is another important difference between clusters of the two
Oosterhoff groups and may indicate something about the evolutionary
status of their RR Lyrae variables.

In Figures 5, 7 and 9, HBR is plotted against mean period. In Figure 5,
we see that clusters with [Fe/H]$<-1.6$ and  blue horizontal
branches (HBR$>0$) belong to the Oosterhoff type II class.
However, having a blue HB does not necessarily guarantee an OoII 
classification because
there are OoI clusters with HBR$>0.5$. On
the other hand, no cluster with HBR$<0$ belongs to the OoII 
class.\footnote{The cluster with HBR=$-0.8$ and $<P(RR0)>, <P_f>=0.617$ is 
Rup 106 which
we have classified as OoI. This anomalous cluster was shown to be extremely
young by Chaboyer et al. (1992).}
According to
Lee et al. (1990), the RR Lyrae variables in OoII clusters
have evolved away from the ZAHB, and as a result, have longer periods and
higher luminosities than the ones in OoI clusters, which are ZAHB
stars. Furthermore, Lee (1990) calculated models illustrating that
for metal poor clusters with HBR$>0.65$,
the bluer the HB morphology of the  cluster, 
the longer the periods should be. To test this
hypothesis, Catelan (1994) made plots of mean periods and HBR using the
cluster data available at the time and found no correlation. It is 
interesting to note that with the new data  
plotted in Figure 5, the cluster with the longest
mean RR0 period  (M2) is the one with the bluest horizontal branch.
Another feature to note in these diagrams, particularly Figure 9,
is that when all the clusters are considered, there is a 
weak correlation between mean period and HBR.

%
%

\section {THE POPULATION II CEPHEIDS}

The period-frequency distrubution for the Cepheids and RV Tauri variables
is shown in the middle panel of Figure 1. The four
RV Tauri variables, plotted
with their pulsation periods which range from about 29 days for V1 in
$\omega$ Cen to 46 days for V17 in M28 seem to form a long period
extension to the Cepheids.
The gap in the period at around 10 days
has an evolutionary origin. 
The longer period variables  are either on a blueward excursion from the AGB
during He-shell flashes or they are on the way towards the white dwarf
cooling region (GS96, Gingold 1976). 
Their periods range from about
12 days (V17 in M14) to 29 days (V21 in M28).

The stars with periods less than 10 days have often been
referred to as BL Her stars, but GS96 pointed out
that this is not an appropriate name because
BL Her itself is not metal deficient. They prefer to call them
AHB1 stars because they have evolved from the blue HB 
and are on the way to the AGB. They cross the instabilty strip at higher
luminosities than the RR Lyrae variables and therefore have longer periods.
It is expected that their periods should increase with time and Wehlau
\& Bohlender (1982) have confirmed this to be the case.
Four of the stars in the short period group
are considered to be anomalous Cepheids. 
These are V19 in NGC 5466 (Zinn \& Dahn 1976) and three $\omega$ Cen stars 
classified by Kaluzny et al. (1997c).
According to WC84, anomalous Cepheids (ACs)
have periods between about 0\day 50 and 3 days, but are
too luminous for their periods. The high luminosities
can be accounted for if they
have larger masses than other population II Cepheids and 
these large masses are thought to occur because they are formed by the 
coalescence of one or more stars.  
Two of the globular cluster ACs in our sample
(Ogle \# 161 in $\omega$ Cen and V19 in NGC 5466)
have periods less than a day.

In Figure 10, we plot [Fe/H] and HBR  versus period for the individual
Cepheids and RV Tauri variables. 
There is no correlation between period and [Fe/H], but most have [Fe/H]
between $-1.25$ and $-1.75$. In fact,
all but one of the Cepheids are found in clusters
that are more metal poor than $-1.25$. 
Horizontal branch morphology 
is an important parameter for determining whether or not a cluster
will have Cepheids.  Generally they occur only in clusters with
blue horizontal branches and this can be seen in Figure 10.
However, we also see in Figure 10 that 
two clusters with HBR$<0$ have Cepheids.
These are NGC 2808 and Pal 3, and in both cases the Cepheids have periods
less than 4 days. Another interesting feature of the diagram is that,
among the Cepheids in
clusters with HBR less than $0.5$, the periods tend to be
longer in clusters with bluer horizontal branches.

It is often assumed that the boundary between
RR Lyrae variables and Cepheids occurs at a period of one day. 
However as we noted in section 2, it is probably shorter. 
WC84 and GS96 have suggested that it may occur at periods as short as
0\day 75 or 0\day 80.
Also, some clusters have bright
variables with periods of about 0\day 5 and light curve
morphology that is significantly different from RR Lyrae variables
with the same period. 
Examples of such stars are V47, V68 and V123 in $\omega$ Cen, V76 in M5 
and V10 in NGC 5897.
Nemec et al. (1994) have suggested that V68 in $\omega$ Cen
might be an anomalous Cepheid or a first-overtone AHB1 variable.
There is also some uncertainty about classification at the longer periods.
V2 in NGC 6712 has been classified as semi-regular by Rosino (1966, 1978)
but Barnes \& Dupuy (1975) considered it to be an RV Tauri variable.
As high precision light curves become available
for more stars, it may  be possible to use light curve morphology, in
addition to color and luminosity,
to distinguish between the different types of pulsating variables.

%
%

\section{THE SR AND RED VARIABLES}

The SR and red variables are the pulsating giant stars that
Rosino (1978) designated as
yellow semi-regular, red semi-regular, irregular or Mira variables  in his 
review of variable stars in globular clusters.  
Whitelock (1986) made a plot of
bolometric magnitude versus the log of the fundamental period for
cluster variables with periods greater than 1 day  and pointed out 
that the relationship for the yellow and red variables might form
an extension to the population II Cepheid P-L relation.

Since not all
authors have designated a specific classification for variables with
longer periods, we have included 
in column 7 of Table 1 
the stars classified as semi-regular or irregular variables and 
all the stars with published
periods greater than 35 days (excluding RV Tauri variables). 
These types of stars have not been well surveyed
in the Galactic globular clusters
because most search programs are designed
to optimize detection and period determination for
variables with periods less than one day.
Thus our sample is probably incomplete.
Periods have been published for some, but not all of these stars because 
of the irregular nature of their variation. 
Many of the `unclassified' variables for which periods have not been
determined (column 3 of Table 1) may belong in this category. 
The period-frequency plot in the bottom panel of Figure 1 shows that
only five globular cluster variables have periods greater than 300 days, but
according to the General
Catalogue of Variable Stars\footnote{The GCVS is
also available at: http://www.sai.msu.su/groups/cluster/gcvs/gcvs/}
(Kholopov 1985a,b, 1987), periods
greater than 300 days are not uncommon among field stars.
Also, in a study of Mira variables in the Galactic bulge, 
Whitelock (1990) reported periods 
ranging up to 720 days with most between 360 and 560 days. 
This is probably a metallicity effect. Mira variables with such long periods 
do not occur in metal poor systems like globular clusters. 
However, in a study of the MACHO data for red giants in the
LMC, Wood et al. (1999) have shown that there are red variables with
periods up to about 1000 days. Because the MACHO observations of the LMC
were made 
throughout the year, for a few years, variations on such a long time scale
could be readily detected.  
Thus, as we have noted above, our
sample of SR/red variables is undoubtedly incomplete.

%
%
\acknowledgements

CMC expresses her appreciation to Marcio Catelan, Pierre Demarque and
Johanna Jurcsik
for the support and encouragement they have given her in the preparation of
this summary.
We would also like to thank Sally MacDonald, David Hogg and James Hogg for 
giving
us the opportunity to examine the bibliographic material prepared by their
mother, Helen Sawyer Hogg. 
Financial support from
the Natural Sciences and Engineering Research Council of Canada (NSERC) is
gratefully acknowledged.
Some of us (A. M., T. R. and J. R.) worked on this project as NSERC summer 
award winners.  
Others (Q. D., T. P., J. W., J. B. and A. R.) have participated 
through the University of Toronto's Research Opportunity Program
for second year undergraduate students.

\newpage

\centerline{\bf{FIGURE CAPTIONS}}

\figcaption[]
{Period-frequency distribution for pulsating variables in Galactic
globular clusters.
The upper panel shows the distribution for stars with periods less than
1 day.  The stars with periods less than 0\day 20 are SX Phe variables.
Most of the others are RR Lyrae variables, but some of the longer period
variables may be anomalous Cepheids or population II Cepheids.
In the middle panel,  the Cepheids and RV Tauri variables are plotted.
The periods plotted for the RV Tauri stars are their pulsation periods. 
Also included
are  two anomalous Cepheids (one in $\omega$ Cen and one in NGC 5466)
with periods less than a day.
In the lower panel, the distribution for
pulsating variables with periods between 50 and 440
days is shown.
\label {Fig. 1}}

\figcaption[]
{Period-frequency distribution for eclipsing binary stars in the Galactic
globular clusters.
\label {Fig. 2}}

\figcaption[]
{Period-frequency distribution for RR Lyrae variables in clusters of
Oosterhoff type I (upper panel) and Oosterhoff type II (lower panel).
\label {Fig. 3}}

\figcaption[]
{Plots of [Fe/H] (from Table 1) versus mean period for the globular cluster
RR0 variables. 
In the upper panel, only clusters with at least 15 RR0 variables
are included and in the lower panels clusters with at least 10 and at least 5,
respectively are included. Filled circles represent clusters that have
red horizontal branches, i.e. HBR=[(B-R)/(B+V+R)]$<0$ and open circles
represent clusters with HBR$>0$. Crosses represent clusters for which HBR has 
not been determined.  
(For $\omega$ Cen, a mean [Fe/H] value, $-1.62$, has been
adopted.)
\label {Fig. 4}}

\figcaption[]
{Plots of HBR (from Table 1) versus mean period for the globular cluster
RR0 variables. In the upper panel, only clusters with at least 15 RR0 
variables
are included and in the lower panels clusters with at least 10 and at least 5,
respectively are included. Filled circles represent clusters that have
[Fe/H]$\ge -1.6$ and open circles
represent clusters with [Fe/H]$<-1.6$. 
\label {Fig. 5}}

\figcaption[]
{Plots of [Fe/H] (from Table 1) versus mean period for the globular cluster
RR1 variables. We include the RR01 stars here as well because their
dominant mode of pulsation is generally the first overtone.
In the upper panel, only clusters with a total of at least 15 RR1 and
RR01 variables
are included and in the lower panels clusters with at least 10 and at least 5,
respectively are included. The symbols are the same as in Figure 4.
\label {Fig. 6}}

\figcaption[]
{Plots of HBR (from Table 1) versus mean period for the globular cluster
RR1 and RR01 
variables. In the upper panel, only clusters with a total of
at least 15 RR1 and RR01 variables
are included and in the lower panels clusters with at least 10 and at least 5,
respectively are included. The symbols are the same as in Figure 5. 
\label {Fig. 7}}

\figcaption[]
{Plots of [Fe/H] (from Table 1) versus mean fundamentalized
period for the globular cluster RR Lyrae variables. 
In the upper panel, only clusters with at least 15 RR Lyrae variables
are included and in the lower panels clusters with at least 10 and at least 5,
respectively are included. The symbols are the same as in Figure 4. 
\label {Fig. 8}}

\figcaption[]
{Plots of HBR (from Table 1) versus mean fundamentalized
period for the globular cluster
RR Lyrae  variables. In the upper panel, only clusters withat least 15 
RR Lyrae variables
are included and in the lower panels clusters with at least 10 and at least 5,
respectively are included. The symbols are the same as in Figure 5. 
\label {Fig. 9}}

\figcaption[]
{Plots of [Fe/H] (upper panel) and HBR (lower panel) versus pulsation period
for the individual Cepheids and RV Tauri variables in GGCs. In the upper 
panel, the symbols are the
same as in Figure 4. 
In the lower panel, the symbols are the same as in Figure 5.
\label {Fig. 10}}

\clearpage
\centerline{\bf{TABLE CAPTIONS}}

{\sc Table} 1. {\sc Numbers and Types of
Variable Stars in Galactic Globular Clusters}

{\sc Table} 2. {\sc Mean Periods of RR Lyrae Variables in Galactic
Globular Clusters}

%

\clearpage

%
%

\begin{deluxetable}{lrccccccl}
\tablecolumns{10}
\tablecaption{Numbers and Types of Variable Stars in Galactic Globular Clusters}
\tablehead{\colhead{Cluster} & \colhead{HBR/[Fe/H]} &
 \colhead{No./} & \colhead{SX} & \colhead{RR} & \colhead{Cep or} 
& \colhead{SR or} & \colhead{Ecl} &\colhead{Ref} \\ 
 && \colhead{(periods)} & \colhead{} & \colhead{}
 &\colhead{RVTau} & \colhead{Red} & \colhead{} 
& \colhead{} \\
\colhead{(1)}& \colhead{(2)}& \colhead{(3)}& \colhead{(4)}& \colhead{(5)}&
\colhead{(6)}& \colhead{(7)}& \colhead{(8)}& \colhead{(9)}
}
\startdata
NGC 104/47Tuc  & -0.99/-0.76 & 50/(37) & 3 &  1  &   0   & 14 & 19 & 1,2,3,4\nl
NGC 288  & 0.98/-1.24  & 10/(10) & 6 &  2  &   0   &  1 & 1 & 5,6\nl
NGC 362  & -0.87/-1.16 & 13/(8)  & 0 &  7  &   0   &  1 & 0 & 7\nl
NGC 1261 & -0.71/-1.35 & 20/(19) & 0 & 18  &   0   &  1 & 0 & 8,9\nl
NGC 1851 & -0.36/-1.22 & 32/(29) & 0 & 29  &   0   &  0 & 0 & 10,11,12,13\nl
NGC 1904/M79 &  0.89/-1.57 &  8/(3)  & 0 &  3  &   0   &  1 & 0 & \nl
NGC 2298 &  0.93/-1.85 &  4/(4)  & 0 &  4  &   0   &  0 & 0 & 14,15\nl
NGC 2419 &  0.86/-2.12 & 41/(33) & 0 & 31  &   1   &  4 & 0 & 16\nl
NGC 2808 & -0.49/-1.15 &  5/(3)  & 0 &  2  &   1   &  2 & 0 & 17\nl
Pal 3    & -0.50/-1.66 & 12/(1)  & 0 &  0  &   1   &  0 & 0 & 18,19,20\nl
NGC 3201 &  0.08/-1.58 & 92/(77) & 0 & 77  &   0   &  0 & 0 & 21,22\nl
Pal 4    & -1.00/-1.48 &  2/(2)  & 0 &  0  &   0   &  2 & 0 & \nl
NGC 4147 &  0.55/-1.83 & 16/(9)  & 0 &  9  &   0   &  0 & 0 & 23\nl
NGC 4372 &  1.00/-2.09 & 16/(13) & 8 &  0  &   2   &  0 & 3 & 24\nl
Rup 106  & -0.82/-1.67 & 16/(16) & 3 & 13  &   0   &  0 & 0 & 25\nl
NGC 4590/M68 &  0.17/-2.06 & 44/(44) & 2 & 42  &   0   &  0 & 0 & 26,27\nl
NGC 4833 &  0.93/-1.79 & 23/(15) & 0 & 14  &   0   &  2 & 0 & 28\nl
NGC 5024/M53      &  0.81/-1.99 & 67/(60) & 0 & 58  &   0   &  2 & 0 & 29,30\nl
NGC 5053 &  0.52/-2.29 & 15/(14) & 5 &  9  &   0   &  0 & 0 & 31\nl
NGC 5139/$\omega$ Cen  & --- /-1.62 & 275/(249) & 34 & 161 & 10 & 
15\tablenotemark{\dag}& 29 & 32,33,34 \nl
 &&&&&&&& 35,36,37\nl 
NGC 5272/M3       &  0.08/-1.57 & 254/(187) & 1 & 182 &  1 &  3 & 0 & 
38,39,40\nl
 &&&&&&&& 41,42,43\nl 
NGC 5286 &  0.80/-1.67 & 24/(13) & 0 & 13  &   0   &  0 & 0 & 44,45,46\nl
NGC 5466 &  0.58/-2.22 & 31/(30) & 6 & 20  &   1   &  0 & 3 & 47,48\nl
 &&&&&&&& 49,50\nl 
NGC 5634 &  --- /-1.82 &  7/(6)  & 0 &  6  &   0   &  0 & 0 & 51\nl
IC 4499  &  0.11/-1.60 & 113/(98) & 1 & 97  &   0  &  0 & 0 & 52,53,54\nl
NGC 5824 &  0.79/-1.85 & 27/(7)  & 0 &  7  &   0   &  0 & 0 & \nl
Pal 5    & -0.40/-1.43 &  5/(5)  & 0 &  5  &   0   &  0 & 0 & \nl
NGC 5897 &  0.86/-1.80 &  9/(9)  & 1 &  7  &   0   &  1 & 0 & 55,56,57\nl
NGC 5904/M5       &  0.31/-1.29 & 158/(135) & 5 & 126 &  2  &  1 & 1 & 
58,59,60,61\nl
 &&&&&&&& 62,63,64\nl 
 &&&&&&&& 65,66,67\nl 
NGC 5927 & -1.00/-0.37 &  9/(5) &  0 & 0  &   0   &  5 & 0 & 68\nl
NGC 5946 &  --- /-1.38  &  6/(2) &  0 & 2  &   0   &  0 & 0 & 69\nl
NGC 5986 &  0.97/-1.58 & 10/(8) &  0 & 8  &   0   &  1 & 0 & 70,71\nl
NGC 6093/M80       &  0.93/-1.75 &  8/(7) &  0 & 6  &   1   &  0 & 0 & 72\nl
NGC 6121/M4       & -0.06/-1.20 & 67/(57) & 4 & 40  &   0   &  1 & 8 & 
73,74,75,76\nl
 &&&&&&&& 77,78,79\nl 
NGC 6101 &  0.84/-1.82 & 15/(0) &  0 &  0  &   0   &  0 & 0 & 80\nl
NGC 6144 &  1.00/-1.73 &  1/(0) &  0 &  0  &   0   &  0 & 0 & 69\nl
NGC 6139 &  0.91/-1.68 & 10/(4) &  0 &  4  &   0   &  0 & 0 & 81\nl
NGC 6171/M107     & -0.73/-1.04 & 23/(22) & 0 & 22  &   0   &  0 & 0 & 82,83\nl
NGC 6205/M13      &  0.97/-1.54 & 29/(18) & 0 &  5  &   5   &  8 & 0 & 
84,85,86\nl
&&&&&&&& 87,87,89\nl 
NGC 6218/M12      &  0.97/-1.48 &  1/(1) &  0 &  0  &   1   &  0 & 0 & 90,91\nl
NGC 6229 &  0.24/-1.43 & 47/(42) & 0 & 38  &   2   &  0 & 2 & 92,93\nl
NGC 6235 &  0.89/-1.40 &  4/(4) &  0 &  3  &   0   &  0 & 1 & 94\nl
NGC 6254/M10       &  0.98/-1.52 &  4/(2) &  0 &  0  &   2   &  1 & 0 & 95\nl
NGC 6266/M62      &  0.32/-1.29 & 89/(74) & 0 & 74  &   0   &  0 & 0 & 96\nl
NGC 6273/M19       &  --- /-1.68 &  5/(5) &  0 &  1  &   4   &  0 & 0 & 97\nl 
NGC 6284 &  --- /-1.32 & 10/(9) &  0 &  6  &   2   &  2 & 0 & 98\nl
NGC 6287 &  0.98/-2.05 &  3/(0) &  0 &  0  &   0   &  0 & 0 & 99\nl
NGC 6293 &  0.90/-1.92 &  5/(4) &  0 &  4  &   0   &  1 & 0 & 100\nl
NGC 6304 & -1.00/-0.59 & 21/(0) &  0 &  0  &   0   &  0 & 0 & 101\nl
NGC 6341/M92    &  0.91/-2.29 & 21/(20) & 2 & 17  &   1   &  0 & 0 & 102,103\nl
NGC 6333/M9        &  0.87/-1.72 & 20/(20) & 0 & 17  &   1   &  1 & 1 & 104\nl
NGC 6356 & -1.00/-0.50 &  8/(3) &  0 & 0  &   0   &  3 & 0 & 68,105\nl
NGC 6352 & -1.00/-0.70 &  3/(1) &  0 &  0  &   0   &  1 & 0 & 68,106,107,\nl
NGC 6366 & -0.97/-0.82 &  1/(1) &  0 & 1  &   0   &  0 & 0 & 108\nl
NGC 6362 & -0.58/-0.95 & 47/(37) & 1 & 35  &   0   &  0 & 1 & 109,110,111\nl
Haute P1 &  --- /-1.50 & 15/(0) &  0 & 0  &   0   &  0 & 0 & \nl
NGC 6380 &  --- /-0.50 &  1/(0) &  0 &  0  &   0   &  0 & 0 & \nl
Terzan 1 &  --- /-0.35 &  4/(0) &  0 &  0  &   0   &  0 & 0 & 112\nl
NGC 6388 &  --- /-0.60 & 29/(10) & 0 & 10  &   0   &  0 & 0 & 68,113,\nl
 &&&&&&&& 114,115\nl 
Pismis 26 & --- /-0.50 & 2/(0) &  0 &  0  &   0   &  0 & 0 & \nl
NGC 6402/M14      &  0.65/-1.39 & 90/(61) & 0 & 54  &   6   &  6 & 0 & 116\nl
NGC 6401 &  --- /-1.12 & 25/(0) &  0 & 0  &   0   &  0 & 0 & 112\nl
NGC 6397 &  0.98/-1.95 &  7/(6) &  2 & 0  &   0   &  0 & 4 & 117,118\nl
NGC 6426 &  0.53/-2.26 & 15/(14) & 0 & 14  &   0   &  0 & 0 & 119,120\nl
Terzan 5 & -1.00/-0.28 &  4/(4) &  0 & 1  &   0   &  2 & 1 & 121,122\nl
NGC 6441 &  --- /-0.53 & 37/(10) & 0 & 10  &   0   &  0 & 0 & 115,123\nl
NGC 6522 &  0.71/-1.44 &  1/(1) &  0 &  1  &   0   &  0 & 0 & 124\nl
NGC 6539 & -1.00/-0.66 &  1/(0) &  0 & 0  &   0   &  0 & 0 & \nl 
NGC 6544 &  1.00/-1.56 &  2/(1) &  0 &  1  &   0   &  0 & 0 & 125\nl
NGC 6541 &  1.00/-1.83 &  1/(0) &  0 &  0  &   0   &  0 & 0 & 126\nl
NGC 6553 & -0.34/-1.00 & 13/(3) &  0 &  2  &   0   &  1 & 0 & \nl  
NGC 6558 &  --- /-1.44 & 15/(9) &  0 & 9  &   0   &  0 & 0 & 127,128\nl
IC 1276/Pal 7    &  --- /-0.73 &  5/(1) &  0 &  1  &   0   &  4 & 0 & \nl
NGC 6569 &  --- /-0.86 & 19/(0) &  0 &  0  &   0   &  0 & 0 & 129\nl
NGC 6584 & -0.15/-1.49 & 46/(42) & 0 & 42  &   0   &  0 & 0 & 130,131\nl
NGC 6624 & -1.00/-0.42 &  4/(0) &  0 &  0  &   0   &  0 & 0 & 132,133\nl
NGC 6626/M28     &  0.90/-1.45 & 20/(15) & 0 & 10  &   3   &  4 & 2 & 
134,135,136\nl
NGC 6638 & -0.30/-0.99 & 61/(19) & 0 & 13  &   0   &  6 & 0 & 137\nl
NGC 6642 &  --- /-1.35 & 18/(16) & 0 & 16  &   0   &  0 & 0 & 138\nl
NGC 6637/M69 & -1.00/-0.71 &  7/(2) &  0 & 0  &   0   &  2 & 0 & \nl
NGC 6652 & -1.00/-0.96 &  5/(0) &  0 &  0  &   0   &  0 & 0 & 139,140\nl
NGC 6656/M22      &  0.91/-1.64 & 36/(24) & 0 & 18  &   1   &  5 & 0 & 
141,142,143\nl
NGC 6681/M70   &  0.96/-1.51 &  3/(2) &  0 & 2  &   0   &  0 & 0 & 144,145\nl
NGC 6712 & -0.64/-1.01 & 19/(14) & 0 & 9  &   0   &  5 & 0 & 146\nl
NGC 6715/M54      &  0.87/-1.59 & 94/(64) & 0 & 61  &   2   &  0 & 1 & 147\nl
NGC 6717 &  0.98/-1.29 &  1/(1) &  0 & 1  &   0   &  0 & 0 & 148\nl
NGC 6723 & -0.08/-1.12 & 32/(28) & 0 & 28  &   0   &  1 & 0 & 68,149,150\nl
NGC 6752 &  1.00/-1.56 & 10/(7) & 3 &  0  &   1   &  0 & 3 & 151,152 \nl   
 &&&&&&&& 153,154\nl 
NGC 6760 & -1.00/-0.52 &  4/(0) &  0 & 0  &   0   &  0 & 0 & \nl
NGC 6779/M56      &  0.98/-1.94 &  7/(4) &  0 & 2  &   2   &  0 & 0 & 155\nl
NGC 6809/M55      &  0.87/-1.81 & 37/(37) & 24 & 13  &   0   &  0 & 0 & 
156,157\nl
NGC 6838/M71     & -1.00/-0.73 & 26/(18) & 3 &  0  &   0   &  2 & 13 & 
88,158\nl
NGC 6864/M75     & -0.42/-1.32 & 17/(6) &  0 & 6  &   0   &  3 & 0 & 159\nl
NGC 6934 &  0.25/-1.54 & 86/(82) & 1 & 78  &   0   &  1 & 2 & 160,161\nl
NGC 6981/M72     &  0.14/-1.40 & 35/(25) & 0 & 25  &   0   &  0 & 0 & 162\nl
NGC 7006 & -0.28/-1.63 & 72/(64) & 0 & 62  &   0   &  1 & 0 & 163,164\nl
NGC 7078/M15      &  0.67/-2.25 & 154/(104) & 2 & 88 &  2   &  0 & 1 &
165,166,167\nl 
 &&&&&&&& 168,169,170\nl 
 &&&&&&&& 171,172\nl 
NGC 7089/M2       &  0.96/-1.62 & 33/(33) & 0 & 29  &   4   &  0 & 0 & 173\nl  
NGC 7099/M30   &  0.89/-2.12 & 13/(3) & 0 &  3  &   0   &  0 & 0 & 112\nl
Pal 12   & -1.00/-0.94 &  3/(0) & 0 & 0  &   0   &  0 & 0 & \nl
Pal 13   & -0.20/-1.65 &  4/(4) & 0 &  4  &   0   &  0 & 0 & \nl
NGC 7492 &  0.81/-1.51 &  4/(4) & 0 &  3  &   1 &  0 & 0 & \nl
\bf TOTAL \rm && 2993(2220) & 117 & 1842 & 60 &  117 & 96 & \nl

\enddata
\end{deluxetable}
\clearpage
{\dag}{$\omega$ Cen contains 5 spotted variables with periods that
range from 3.3 days to 34 days.  It also contains 2 possible ellipsoidal 
variables.  Since variables of these types have only been identified in 
$\omega$ Cen, they are included in this column, but they are not plotted
in Figure 1. There are only 3 SR/red variables in $\omega$ Cen.

References. ---
(1) Fox (1982); (2) Carney et al. (1993); (3) Edmonds et al. (1996);
(4) Kaluzny et al. (1998b);
(5) Kaluzny (1996); (6) Kaluzny et al. (1997a); (7) Lloyd Evans (1978);
(8) Wehlau \& Demers (1977); (9) Wehlau et al. (1977); (10) Liller (1975);
(11) Wehlau et al. (1978); (12) Wehlau et al. (1982); (13) Walker (1998);
(14) Liller (1976); (15) Clement et al. (1995a); (16) Pinto \& Rosino (1977);
(17) Clement \& Hazen (1989); (18) Gratton \& Ortolani (1984); (19) Borissova
et al. (1998); (20) Borissova et al. (2000); (21) Lee (1977b); (22) Cacciari
(1984a,b); (23) Clement (2000); (24) Kaluzny \& Krzemi\'nski (1993); 
(25) Kaluzny et al. (1995); (26) Clement et al. (1993); (27) Walker (1994); 
(28) Demers \&
Wehlau (1977); (29) Goranskij (1976); (30) Kopacki (2000); (31) Nemec et al.
(1995a, 1995b); (32) Butler et al. (1978); (33) Fourcade et al. (1978);
(34) Liller (1978); (35) Niss et al. (1978); (36) Jorgensen \& Hansen (1984);
(37) Kaluzny et al. (1996, 1997b,c); (38) Kholopov (1974, 1977); (39)
Meinunger (1980); (40) Kadla \& Geraschenko (1982); (41) Kaluzny et al. 
(1998a);
(42) Corwin et al. (1999a); (43) Bakos et al. (2000); (44)
Fourcade et al. (1975); (45) Liller \& Lichten (1978a); (46) Gerashchenko et al.
(1997); (47) Buonanno et al. (1984); (48) Mateo et al. (1990); (49) Nemec \&
Mateo (1990); (50) Corwin et al. (1999b); (51) Liller \& Sawyer Hogg (1976);
(52) Clement et al. (1979); (53) Clement et al.  (1986); (54) Walker \& Nemec 
(1996); (55) Wehlau (1990); (56) Wehlau et al. (1996); (57) Clement \& Rowe 
(2001); (58) Kadla et al. (1987); (59) Kravtsov (1988, 1991, 1992);
(60) Sandquist et al. (1996); (61) Brocato et al. (1996); (62) Reid (1996);
(63) Yan \& Reid (1996); (64) Drissen \& Shara (1998); (65) Olech et al.
(1999b); (66) Kaluzny et al. (1999, 2000); (67) Caputo et al. (1999);
(68) Lloyd Evans \& Menzies (1977); (69) Liller (1983a); (70) Liller \&
Lichten (1978b); (71) Alves et al. (2001);
(72) Wehlau et al. (1990); (73) Sujarkova \& Shugarov (1981);
(74) Lee (1977a); (75) Yao et al. (1981, 1988); 
(76) Yao (1986, 1987, 1991, 1993);
(77) Cudworth \& Rees (1990); (78) Shokin \& Samus (1996); (79) Kaluzny et al.
(1997d); (80) Liller (1981); (81) Hazen (1991); (82)
Clement \& Sawyer Hogg (1977a); (83) Clement \& Shelton (1997); (84) Russev
\& Russeva (1979a,b); (85) Russeva \& Russev (1980,, 1983) (86) Russeva et al. 
(1982)  (87) Kadla et al. (1980);
(88) Welty (1985); (89) Osborn (2000a,b); (90) Clement et al. (1988); (91)
Malakhova et al. (1997a); (92) Carney et al. (1991); (93) Borrisova et al. 
(1997, 2001); (94) Liller (1977); (95) Clement et al. (1985); (96) Malikhova
et al. (1997b); (97) Clement \& Sawyer Hogg (1978); (98) Clement et al. (1980);
(99) Stetson \& West (1994); (100) Clement et al. (1982); (101) Hartwick et al.
(1981); (102) Kadla et al. (1983); (103) Kopacki (2001); (104) Clement et al.
(1984), Clement \& 
Shelton (1996); (105) Clement \& Sawyer Hogg (1977b); 
(106) Whitelock (1986);
(107) Hesser (1980); (108) Harris (1993); (109) Clement et al. (1995b);
(110) Mazur et al. (1999); (111) Olech et al. (2001); (112) Terzan \&
Rutily (1975); (113) Hazen \& Hesser (1986); (114) Silbermann et al. (1994);
(115) Pritzl et al. (2000); (116) Wehlau \& Froelich (1994); (117) Kaluzny
(1997); (118) Cool et al. (1998); (119) Clement \& Nemec (1990); 
(120) Papadakis et al. (2000); (121) Spinrad et al. (1974); (122) Edmonds
et al. (2001); (123) Layden et al. (1999); (124) Walker \& Mack (1986);
(125) Hazen (1993a); (126) Hazen (1994); (127) Hazen (1996); (128) Blanco \&
Blanco (1997); (129) Hazen-Liller (1985); (130) Millis \& Liller (1980); 
(131) Samus et al. (1995); (132) Liller \& Liller (1976); (133) Deutsch et al.
(1999); (134) Wehlau \& Sawyer Hogg (1982); (135) Wehlau \& Butterworth (1991);
(136) Rees \& Cudworth (1991); (137) Rutily \& Terzan (1977); (138) Hazen 
(1993b); (139) Hazen (1989); (140) Deutsch et al. (2000); (141) Wehlau \&
Sawyer Hogg (1977, 1978); (142) Marinchev (1983); (143) Kravtsov et al. (1994);
(144) Liller (1983b); (145) Kadla et al. (1996); (146) Cudworth (1988);
(147) Layden \& Sarajedini (2000); (148) Goranskij (1978); (149) Menzies
(1974); (150) Kovacs et al. (1986); (151) Cannon \& Stobie (1973); 
(152) Lee (1974);
(153) Thompson et al. (1999); (154) Bailyn et al. (1996); (155) Wehlau \&
Sawyer Hogg (1985); (156) Olech et al. (1999a); (157) Pych et al. (2001);
(158) Park \& Nemec (2000); (159) Pinto et al. (1982); (160) Sawyer Hogg \&
Wehlau (1980); (161) Kaluzny et al. (2001); (162) Kadla et al. (1995);
(163) Pinto \& Rosino (1973); (164) Wehlau et al. (1999); (165) Chu (1977);
(166) Chu et al. (1984); (167) Kadla et al. (1984); (168) Geffert et al.
(1989); (169) Yao (1990), Yao \& Qin (1993); (170) Silbermann \& Smith 
(1995); (171)
Butler et al. (1998); (172) Jeon et al. (2000); (173) Lee \& Carney (1999a).

\clearpage
\begin{deluxetable}{lccccccccc}
\tablecaption{Mean Periods of RR Lyrae Variables in Galactic Globular Clusters}
\tablehead{
& \colhead{\# of} &\colhead{$<P>$} & \colhead{\# of} & \colhead{$<P>$} & 
\colhead{\# of} & \colhead{$<P>$} & 
\colhead{\# of} & \colhead{$<P>$} & \colhead{$<P_{f}>$} \\ 
 \colhead{Cluster} & \colhead{RR0} & \colhead{(RR0)} & \colhead{RR1} 
& \colhead{(RR1)} & \colhead{RR2} & \colhead{(RR2)} &
  \colhead{RR01}& \colhead{(RR01)} & \\ 
\colhead{(1)} & \colhead{(2)} & \colhead{(3)} & \colhead{(4)} & \colhead{(5)} &
\colhead{(6)} & \colhead{(7)} & \colhead{(8)} & \colhead{(9)} & \colhead{(10)} 
}
\startdata
NGC 104  &   1 &  0.737 & 0 & &  0 & &  0  & & 0.737 \nl 
NGC 288  &   1 &  0.678 & 1 &  0.430 &  0 & &  0  &  & 0.628 \nl   
NGC 362  &   7 &  0.542 &  0 & &  0 & &  0 & & 0.542 \nl 
NGC 1261 &  13 &  0.555 &  5 & 0.328 &  0 & &  0  & & 0.523 \nl  
NGC 1851 &  21 & 0.571  &  7 & 0.317 &   1  & 0.266 & 0  & & 0.531 \nl 
NGC 1904 &   2 &  0.685 &  1 & 0.335 &   0 & &  0 & &  0.607 \nl 
NGC 2298 &   1 &  0.640 &  3 & 0.394 &  0  & &  0  & & 0.557   \nl 
NGC 2419 &  24 &  0.655 &  6 & 0.380 &  0  & &  1  & 0.407 & 0.624 \nl
NGC 2808 &   1 &  0.539 &  0 & &  1 & 0.306 & 0 & & 0.526 \nl 
NGC 3201 &  72 &  0.554 &  3 & 0.338 &  2  &  0.274& 0 & & 0.547  \nl 
NGC 4147 &   4 &  0.531 &  5 & 0.339 & 0 & &  0 & & 0.489   \nl
Rup 106  &  13 &  0.617 &  0 & &  0 & &  0 & & 0.617 \nl 
M68     &  13 &  0.613 & 17 & 0.390 & 0 & & 12 & 0.397 & 0.554 \nl
NGC 4833 &   7 &  0.708 &  7 & 0.380 & 0 & &  0 & & 0.609 \nl  
M53     &  29 &  0.649 & 27 & 0.351 & 2 & 0.309 &  0 & & 0.562  \nl
NGC 5053 &   5 &  0.672 &  4 & 0.354 & 0 & &  0 & & 0.585 \nl  
$\omega$ Cen&  76 &  0.651 & 59 & 0.387 & 26  & 0.306 & 0 & & 0.581   \nl 
M3       & 145 &  0.555 & 27 & 0.345 &  5  & 0.286 & 5 & 0.358 & 0.537 \nl
NGC 5286 &   8 &  0.614 & 5  & 0.344 &  0  & &  0  & & 0.556  \nl 
NGC 5466 &  13 &  0.646 & 5  & 0.382 &  2  & 0.262 & 0 & & 0.592  \nl 
NGC 5634 &   3 &  0.621 & 3  & 0.379 &  0 & & 0 & & 0.565 \nl 
IC 4499  &  63 &  0.580 & 14 & 0.354 & 3  & 0.284 & 17  & 0.358 & 0.544 \nl
NGC 5824 &   7 &  0.624 & 0 & &  0 & &  0 & & 0.624 \nl      
Pal 5    &   0 & &  3  & 0.316 & 2  & 0.273 & 0 & & 0.438  \nl 
NGC 5897 &   3 &  0.828 &  2 & 0.437 & 2 & 0.345 & 0 & & 0.688 \nl 
M5       &  91 &  0.551 & 34 & 0.320 & 1  &  0.265 & 0 & & 0.517  \nl 
NGC 5946 &   2 &  0.651 & 0 & &  0 & &  0 & & 0.651  \nl 
NGC 5986 &   7 &  0.652 & 1  &  0.328 & 0 & &  0 & & 0.626 \nl  
M80   &  4  &  0.651 &  2 &  0.366 & 0 & & 0  & & 0.598 \nl 
M4       &  31 &  0.533 & 7  & 0.302  & 2  & 0.250 & 0 & & 0.505   \nl 
NGC 6139 &   3 &  0.662 & 1  & 0.417 & 0 & &  0 & & 0.636  \nl  
M107        &  15 &  0.538 & 7  & 0.287 & 0 & &  0 & & 0.490  \nl  
M13      &   1 &  0.750 & 3  & 0.358 & 1  & 0.313 & 0 & & 0.543    \nl 
NGC 6229 &  30 &  0.553 & 7  & 0.330 & 1  & 0.264 & 0 & & 0.529    \nl 
NGC 6235 &   2 &  0.600 & 1  & 0.352 & 0  & &  0 & & 0.557 \nl   
M62   &  62 &  0.544 & 11 & 0.304 & 1  & 0.248 & 0 & & 0.522 \nl 
M19     &   1 &  0.507 & 0 & &  0 & &  0 & & 0.507  \nl  
NGC 6284 &   6 &  0.588 & 0 & &  0 & &  0 & & 0.588  \nl 
NGC 6293 &   2 &  0.600 & 2  & 0.353 & 0 & &  0 & & 0.537    \nl  
M92  &  11 &  0.630 & 5  & 0.364 & 1 & 0.313 &  0 & & 0.582    \nl 
M9  &   8 &  0.638 & 9  & 0.342 & 0 & &  0 & & 0.543   \nl 
NGC 6366 &   1 &  0.513 & 0 & &  0 & &  0 & & 0.513  \nl
NGC 6362 &  18 &  0.547 & 15  & 0.299 & 2  & 0.251 & 0 & & 0.477 \nl
NGC 6388 &   5 &  0.712 & 5  &  0.332 & 0  & & 0  & & 0.579 \nl 
M14      &  39 &  0.564 & 10  & 0.335 & 5  &  0.320 & 0 & & 0.540     \nl 
NGC 6426 &   9 &  0.704 & 4  & 0.332  & 0  & &  1  & 0.408 & 0.619 \nl
Terzan 5 &   1 &  0.60  & 0 & & 0 & & 0 & & 0.60 \nl
NGC 6441 &  10 &  0.768 & 0 & & 0 & & 0 & & 0.768 \nl
NGC 6522 &   1 &  0.564 & 0 & &  0 & &  0 & & 0.564  \nl   
NGC 6544 &   1 &  0.57  & 0 & &  0 & &  0 & & 0.570  \nl   
NGC 6553 &   2 &  0.526 & 0 & &  0 & &  0 & & 0.526  \nl   
NGC 6558 &   6 &  0.556 & 3 & 0.345 & 0 & &  0  & & 0.525  \nl   
Pal 7    &   1 &  0.548 & 0 & &  0 & &  0 & & 0.548  \nl   
NGC 6584 &  34 &  0.560 & 8  & 0.304 & 0 & &  0 & & 0.531 \nl
M28      &   8 &  0.577 & 2  & 0.312 & 0 & &  0 & & 0.545    \nl   
NGC 6638 &   1 &  0.666 & 11 & 0.307 & 1  & 0.308 & 0  & & 0.439 \nl  
NGC 6642 &  10 &  0.544 & 6  & 0.322 & 0 & &  0 & & 0.502 \nl  
M22  &  10 &  0.632 & 8  & 0.361 & 0 & &  0 & & 0.567 \nl
M70   &   1 &  0.564 & 1  & 0.402 & 0 & &  0 & & 0.552 \nl   
NGC 6712 &   7 &  0.557 & 2  & 0.338 & 0 & &  0 & & 0.534 \nl
M54   &  55 &  0.579 & 6  & 0.342 & 0 & &  0 & & 0.568 \nl
NGC 6717 &   1 &  0.575 & 0  & & 0  & & 0  & & 0.575  \nl
NGC 6723 &  23 &  0.541 & 4  & 0.292 & 1  & 0.288 & 0 & & 0.517 \nl
M56   &   1 &  0.906 & 1 & 0.423 &  0 & &  0 & & 0.736 \nl
M55      &   4 &  0.662 & 5  & 0.391 & 4  & 0.321 & 0 & & 0.571    \nl 
M75   &   3 &  0.531 & 3  & 0.271 & 0 & &  0 & & 0.447    \nl
NGC 6934 &  68 &  0.574 & 9  & 0.308 & 1 & 0.247 &  0 & & 0.553   \nl
M72  &  24 &  0.547 & 1  & 0.353 & 0 & &  0 & & 0.544    \nl
NGC 7006 &  53 &  0.569 & 9  & 0.329 & 0 & &  0 & & 0.550   \nl
M15      &  39 &  0.637 & 30 & 0.337 & 2  & 0.290 & 17 & 0.401 & 0.551 \nl
M2       &  17 &  0.725 & 9  & 0.345 & 3  & 0.298 & 0  & & 0.621 \nl
M30 &   3 &  0.698 & 0 & &  0 & &  0 & & 0.698 \nl  
Pal 13   &   4 &  0.572 & 0 & &  0 & &  0 & & 0.572  \nl 
NGC 7492 &   1 &  0.805 & 1  & 0.292 & 1  & 0.280 & 0 & & 0.556 \nl
\bf TOTAL \rm & 1269 & 0.585 & 447 & 0.345 & 73 & 0.296 & 53 & 0.383 & 0.550\nl

\enddata
\end{deluxetable}

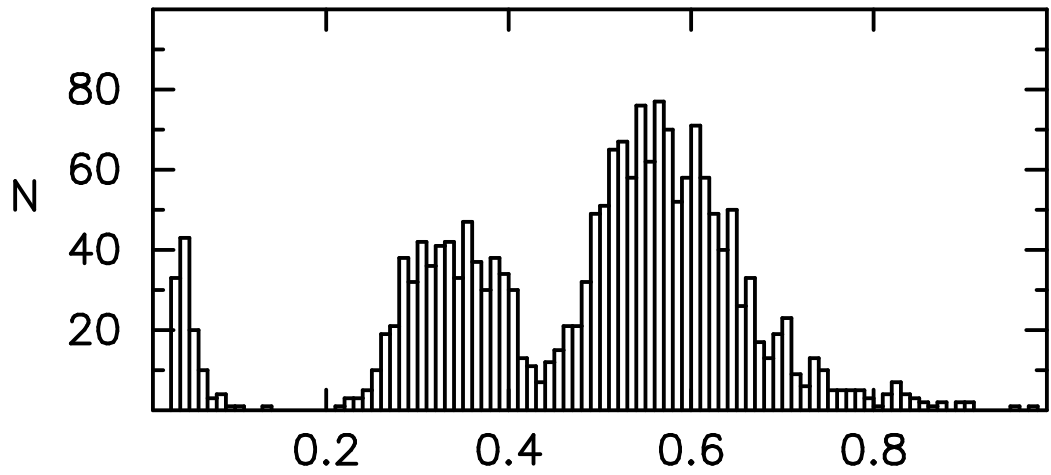
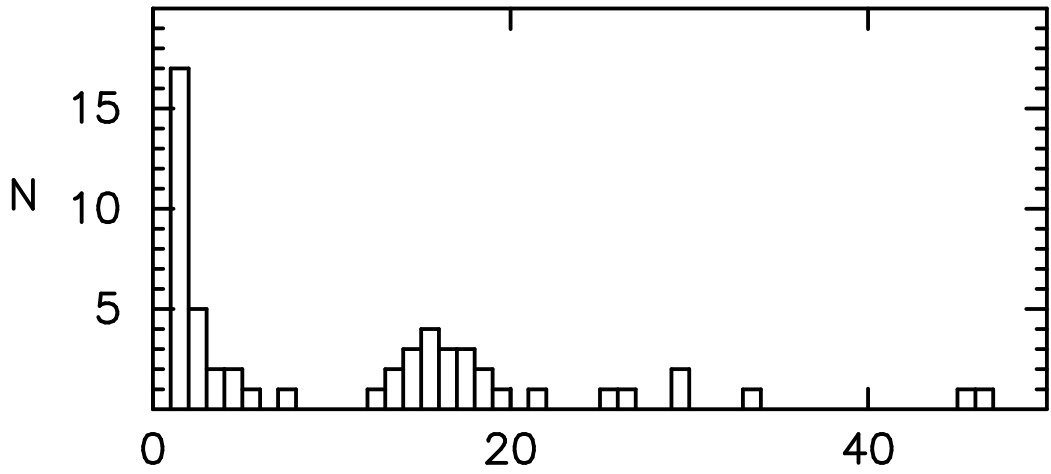
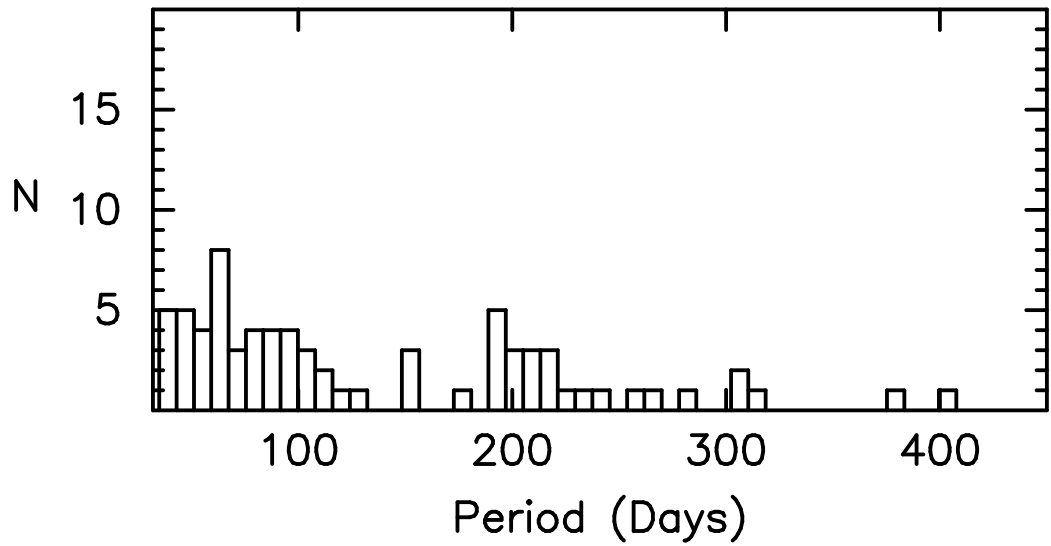
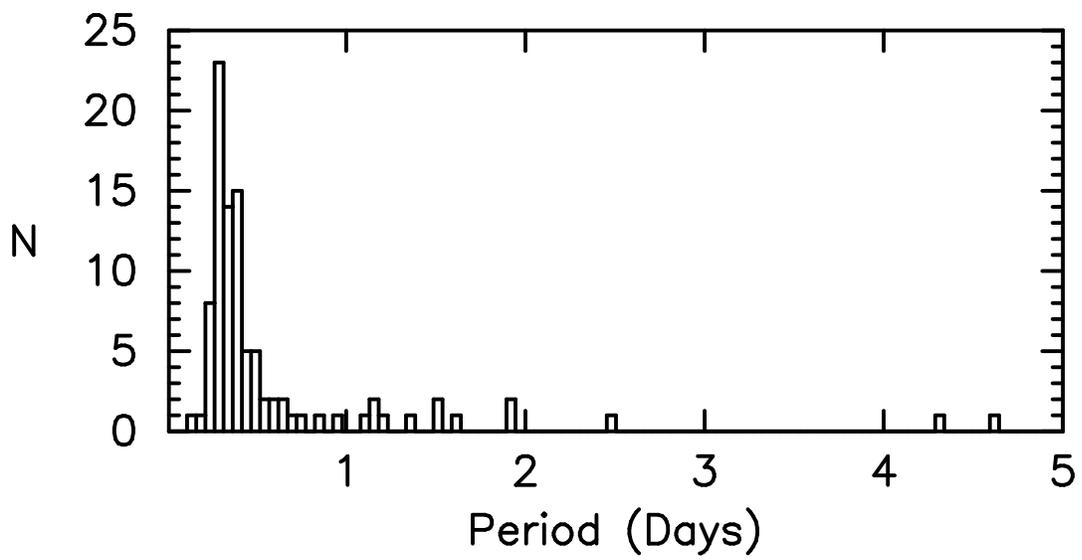
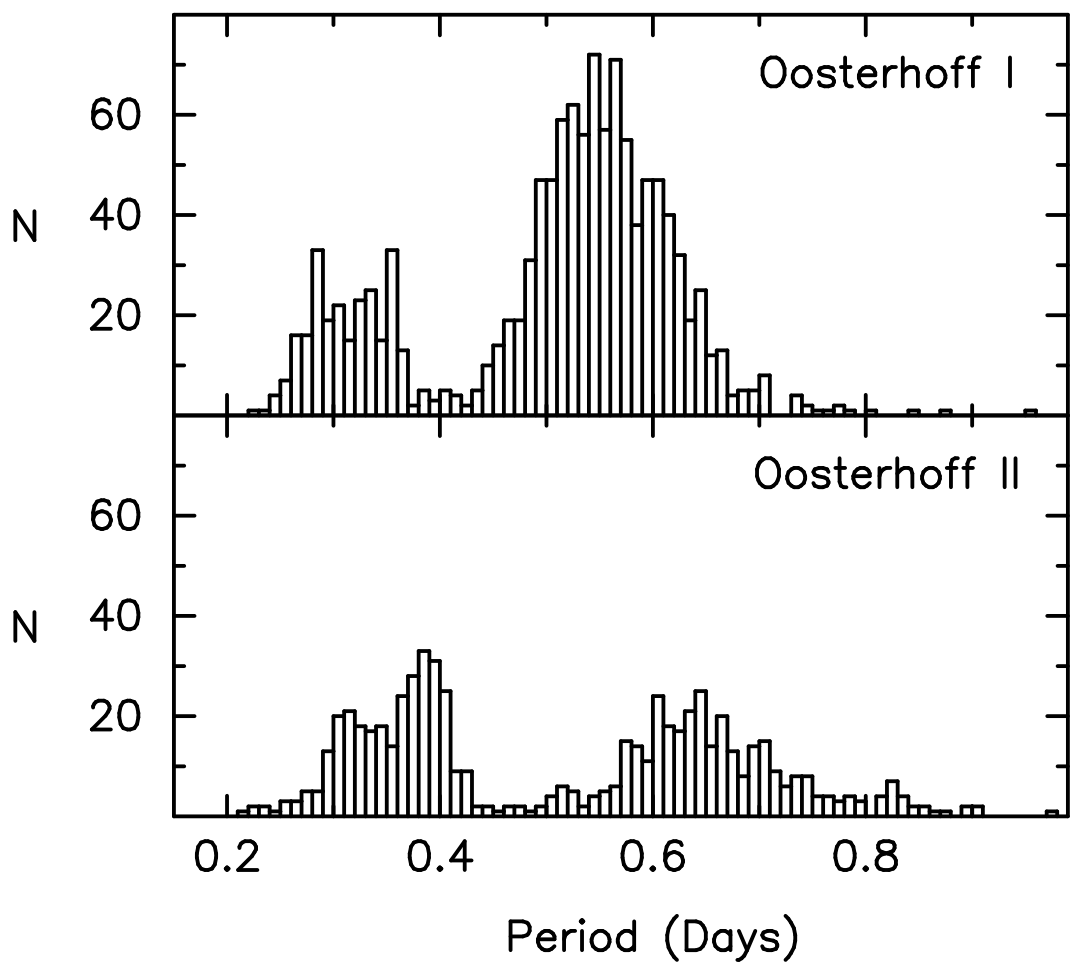
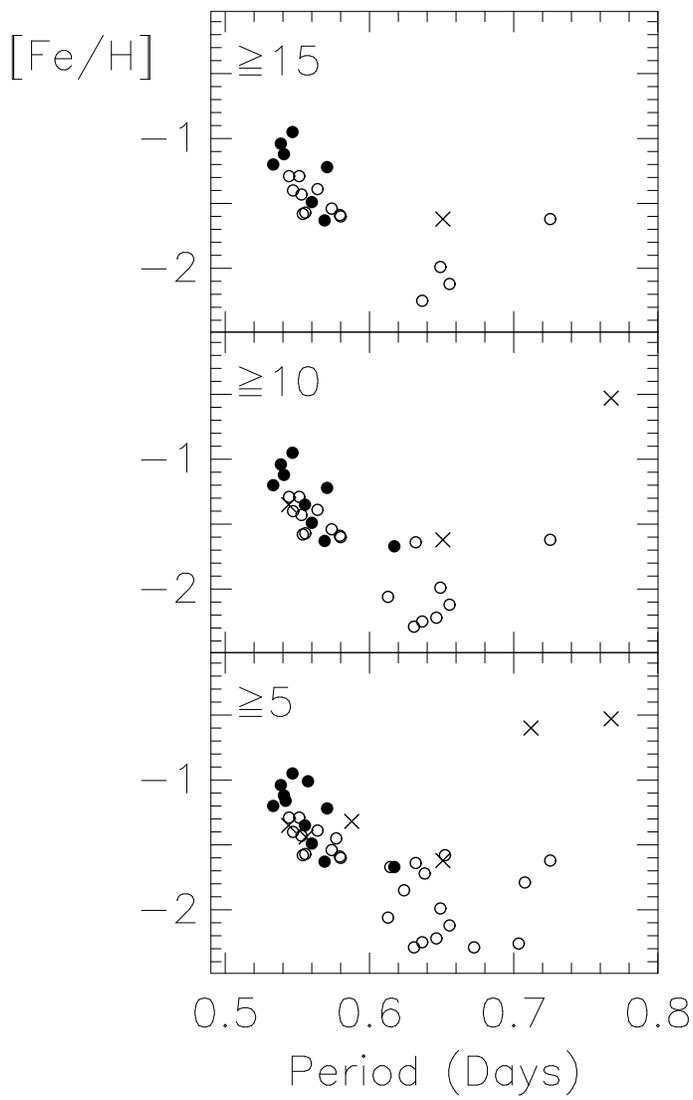
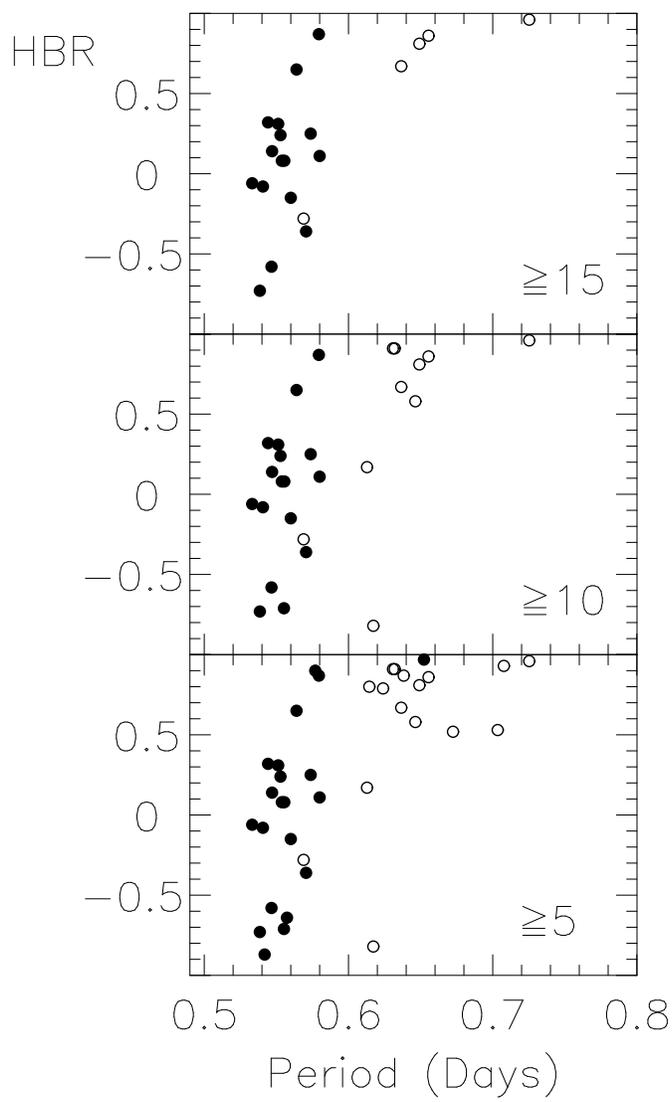
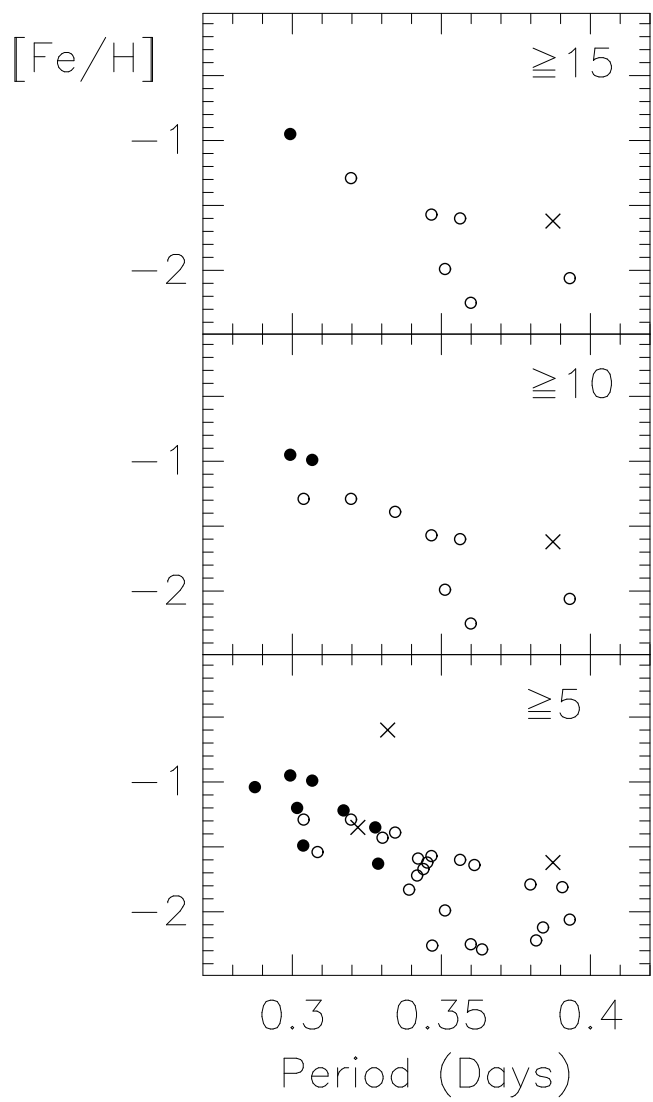
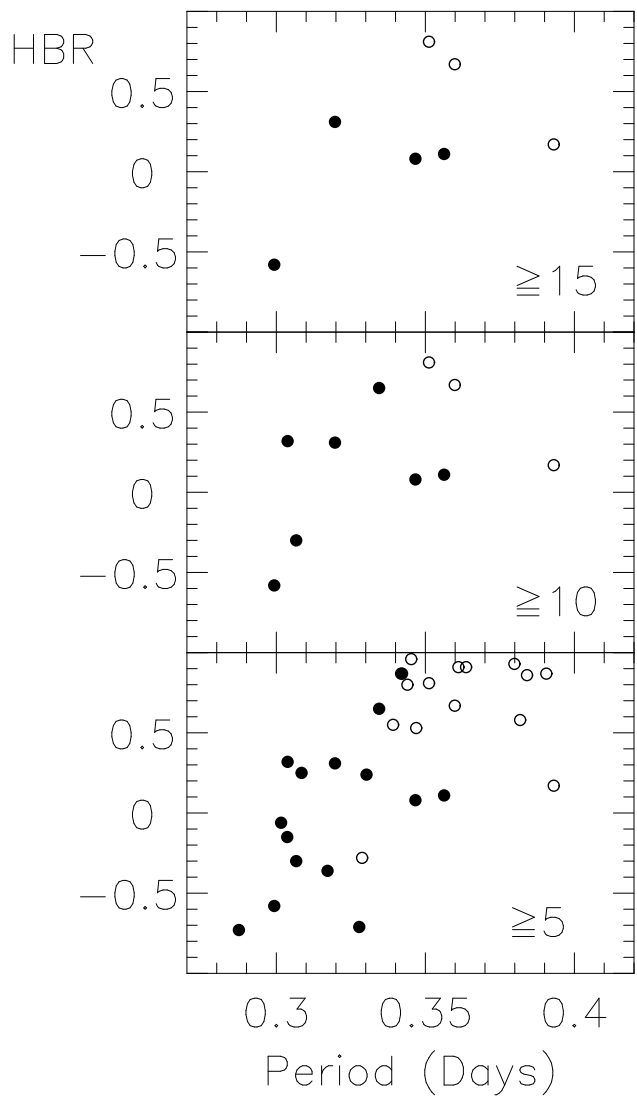
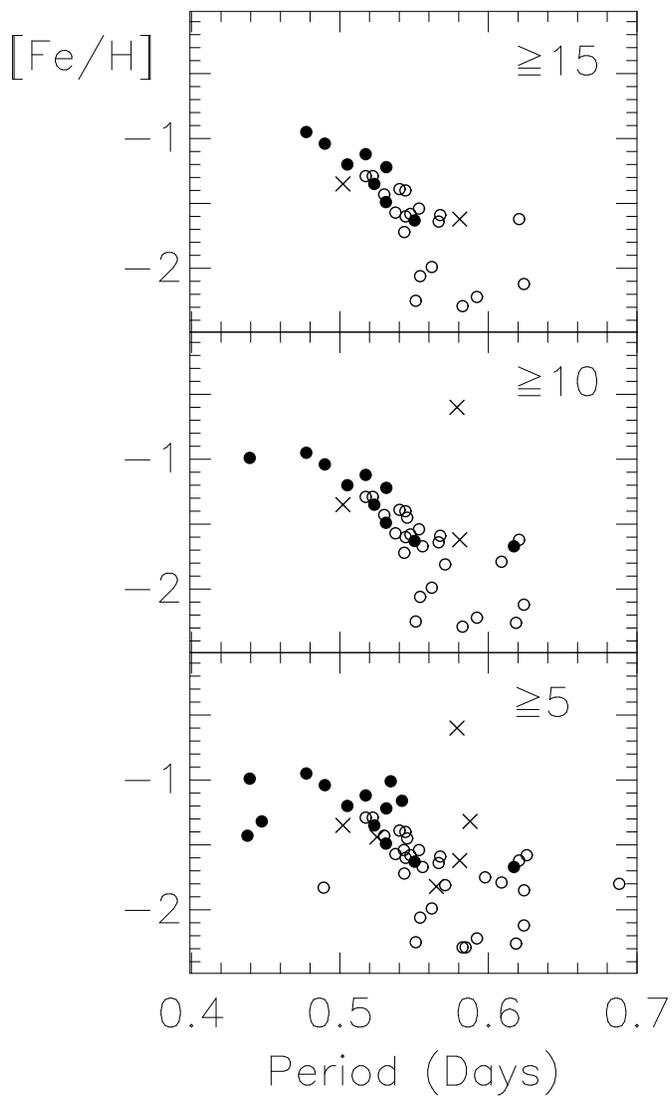
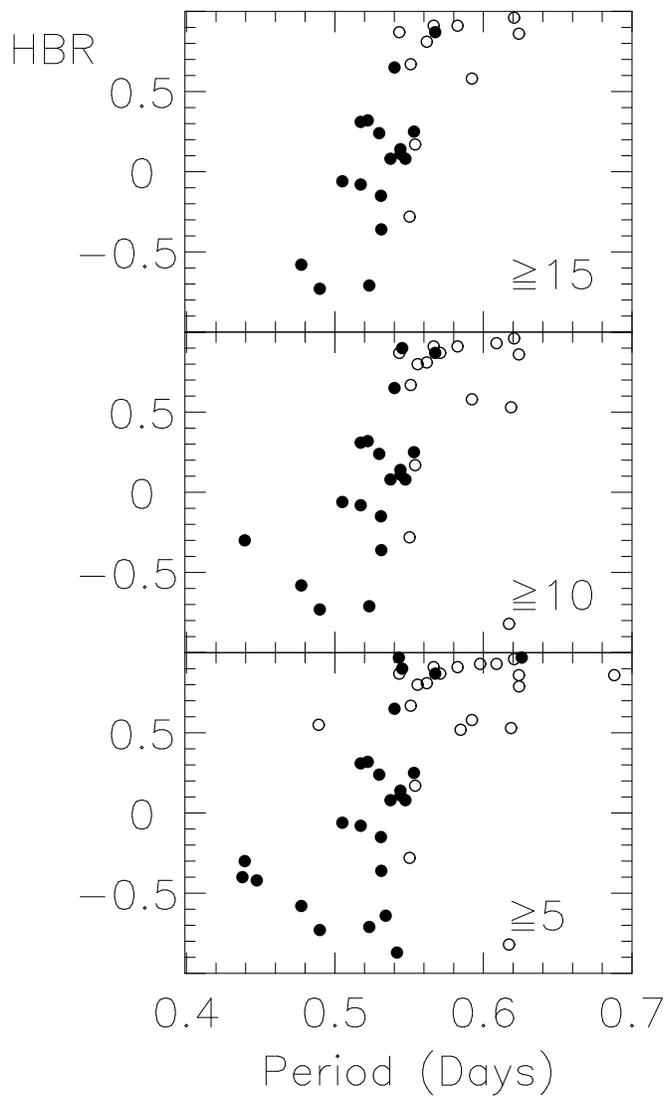
\end{document}